\documentclass[prl,twocolumn,preprintnumbers,amsmath,amssymb]{revtex4}

\usepackage{graphicx}
\usepackage{ifthen}
\usepackage{ifpdf}
\usepackage{color}

\usepackage{latexsym}
\usepackage{amsmath}
\usepackage{amssymb}
\usepackage{bm}
\usepackage{ulem}

\usepackage{color}

\usepackage{ifpdf}

\ifpdf

\usepackage[pdftex]{hyperref}
\else
\usepackage[hypertex,ps2pdf,dvips,colorlinks,urlcolor=blue,citecolor=blue]{hyperref}
\fi

\def\be{\begin{equation}}
\def\ee{\end{equation}}

\definecolor{Blue}{rgb}{0.00, 0.00, 1.00}
\definecolor{Red}{rgb}{1.00, 0.00, 0.00}

\begin{document}

\title{Stationary entropies after a quench from excited states in the Ising chain}

\author{ M\'arton Kormos$^{1}$, Leda Bucciantini$^{2}$, Pasquale Calabrese$^{2}$}

\affiliation{ $^1$
MTA-BME ÒMomentumÓ Statistical Field Theory Research Group, 1111 Budapest, Budafoki \'ut 8, Hungary}
\affiliation{ $^2$  Dipartimento di Fisica dell'Universit\`a di Pisa and INFN, 56127 Pisa, Italy}

\begin{abstract}

We consider the asymptotic state after a sudden quench of the magnetic field in the transverse field  quantum  Ising chain starting from excited states of the pre-quench Hamiltonian. We compute the thermodynamic entropies of the generalised Gibbs and the diagonal ensembles  and we find that 
the generalised Gibbs entropy is always twice the diagonal one. 
We show that particular care should be taken in extracting the thermodynamic limit since different averages of equivalent microstates give different results for the entropies.

\end{abstract}

\maketitle

Entropy is a fundamental concept of statistical mechanics and represents the main bridge between the microscopic 
description of nature and thermodynamics.  
Indeed, a generic isolated {\it classical} system evolves in a way to maximise its entropy reaching the microcanonical ensemble after a long time. 
In the quantum world the situation is more complicated: an isolated system evolves unitarily, so if the system is initially prepared in a pure state it will always remain pure with strictly zero entropy, and cannot be described asymptotically by a statistical ensemble with positive entropy.
The definition of a stationary entropy for non-equilibrium quantum systems is then a complicated matter which attracted renewed interest after the cold atom experimental realisation \cite{exp,revq}  of isolated out of equilibrium quantum systems, in particular of the so called quantum quenches \cite{cc-06,cc-07}, in which a  parameter of the system is changed abruptly.

Two main roads have been followed to define a non-equilibrium stationary entropy after a quench. 
The first is to look at the system in its entirety and define the entropy in a specific basis, such as for the so-called diagonal entropy \cite{pol-11}.
The second road is to consider subsystems of the whole system which are not isolated and therefore are described by a reduced density matrix that may be equivalent to a statistical ensemble. 
The two roads have both their own advantages and disadvantages. 
Indeed, considering only a subsystem is probably more appropriate from a fundamental perspective because 
taking first the thermodynamic (TD) and then the large time limit (see e.g. \cite{bs-08,cdeo-08,cef,cef-i,cef-ii}), 
it is possible that the reduced density matrix exhibits truly stationary behaviour which is impossible for the entire system. 
Conversely, a global definition of entropy is surely more suitable and manageable for finite systems and numerical simulations
\cite{spr-11,sr-10,v-e}.

The possible connections and relations between these two apparently unrelated sets of (stationary) entropies are then very important. Some explicit calculations 
for integrable systems in quenches from the ground state of a pre-quench Hamiltonian show that the 
diagonal entropy is exactly half of the subystem entropy \cite{g-13,ckc-13}, reflecting the fact that the diagonal ensemble
contains much more information than the one needed to describe the local observables.

However, the previous studies focused on the evolution starting from the ground state of a given pre-quench Hamiltonian.
Starting from an excited state makes the situation more complicated. 
Indeed, while the ground state is usually unique (or with low degeneracy), there are many excited states which share the same macroscopical properties and some kind of average among them should be introduced in order to extract the TD limit. 
It is then a very relevant question if and how the average changes the expectation values of the entropies and if it has the same effect on the two kinds of entropies. 

In order to shed some light to this problem we consider here the simplest exactly solvable model, 
the transverse field Ising chain with Hamiltonian 
\begin{equation}
\label{h_Ising chain}
H(h)=-\frac{1}{2} \sum_{j=1}^{N}{[\sigma^x_j\sigma^x_{j+1}+h\sigma_j^z]}\,,
\end{equation} 
where $\sigma^\alpha_j$, $\alpha=x, y,z$ are the Pauli matrices at site $j$ of the chain of length $N$, $h$ is the transverse field and 
periodic boundary conditions are imposed. 
The model can be mapped to spinless free fermions through the Jordan--Wigner transformation, 
and diagonalised by a further Bogoliubov transformation in momentum space (see e.g. \cite{sach-book} for details), yielding
\begin{equation}\label{H_diag_preq}
H(h)=\sum_k{\epsilon_k\left(b^\dagger_k  b_k-\frac12\right)}\,,
\end{equation}
where $b_k$, $b_k^\dagger$  are the annihilation and creation operator for the fermionic quasi-particles 
(i.e. the Bogoliubov modes) of momentum (wave number) $k$, satisfying canonical anticommutation relations $\{ b_k, b^\dagger_{k'}\}=\delta_{kk'}$, 
with one-particle dispersion relation 
\begin{equation}
\epsilon_k=\sqrt{\left(h-\cos p_k \right)^2+\sin^2\left(p_k\right)}\,, \quad\quad p_k= \frac{2\pi k}{N}\,.
\end{equation}

We are interested in quenches when at $t=0$  the transverse magnetic field is suddenly switched from $h_0$ to $h$. We consider as initial state an excited state of $H(h_0)$ which is neither an eigenstate nor a finite superposition of eigenstates of $H(h)$.
We restrict our attention to the class of excited states which can be written acting on the ground state of 
$H(h_0)$ with an arbitrary number of pre-quench creation operators $b'^\dagger_k$  for the modes $k$ 
(here and below primed quantities will denote pre-quench ones). 

Given that the pre-quench ground state 
$|0\rangle'$ is defined by $b'_k|0\rangle'=0 \quad\forall k $, in the Fock basis 
the initial state can be written as 
\be
|\Psi_0\rangle \equiv\prod_{k} (b'^\dag_k)^{m_k}|0\rangle'\,.
\label{ES}
\ee
It has energy $E_{m_k}-E_{GS}=\sum_{k} m_k \epsilon_k$, where 
 $m_k=0,1$ is a characteristic function of the state representing  the pre-quench fermionic occupation number, 
i.e. $m_k=1$ if the mode $k$ is occupied and $m_k=0$ if it is not.  
Since $k$ can assume $N$ possible values, there are $2^N$ linearly independent eigenstates of this form which 
consequently form a basis. 
While in a finite system the characteristic function takes only the values $0$ and $1$, 
in the TD limit it becomes an arbitrary function $m(p)$
of the continuous momentum variable $p\in [-\pi,\pi]$ with the 
restriction to be in the interval $[0,1]$ (see for instance \cite{afc-09,bkc-14}). 
It is evident that there are many micro-states described in the TD limit by the same function $m(p)$.
All TD quantities can depend specifically on the way the average over those micro-states  
is performed and what happens to the entropy is one of the main points we aim to clarify in this manuscript.

\textbf{Relaxation to thermodynamic ensembles}.
Denoting the time dependent state by $|\Psi(t)\rangle=e^{-i H(h)t} |\Psi_0\rangle$, the density matrix of the entire system is
\be
\rho(t)=|\Psi(t)\rangle\langle\Psi(t)|\,,
\ee
and the reduced density matrix of a subsystem consisting of a block $A$ of $\ell$ contiguous spins is
\be
\rho_A(t)={\rm Tr}_{\bar{A}} \,\rho(t)\,,
\ee
where $\bar{A}$ is the complement of $A$. The importance of $\rho_A$ stems from the fact that it generically relaxes to a stationary 
state described by some statistical ensemble, while the full density matrix $\rho(t)$ always describes a pure state with zero entropy. 
The von Neumann entropy of $\rho_A(t),$
\be
S_A(t)\equiv - \mathrm{Tr}\rho_{A}(t)\ln\rho_{A}(t)\,,
\ee
is the well known entanglement entropy which has also been studied for quenches in the Ising chain \cite{cc-05,fc-08,bkc-14}.

Following Refs. \cite{bs-08,cdeo-08,cef-ii} it is usually said that a system reaches a stationary state if a long time limit 
of the reduced density matrix
\be
\lim_{t\to\infty}\rho_A(t) =\rho_A(\infty)
\ee
exists. 
This stationary state is said to be described by a statistical  ensemble with density matrix $\rho_E$
if the reduced density matrix of the latter restricted to any finite subsystem  $A$  equals $\rho_A(\infty)$, i.e. if 
for $\rho_{A, E}\equiv {\rm Tr}_{\bar{A}}(\rho_E)$ 
\be
\rho_A(\infty)=\rho_{A, E}\,.
\ee
In particular, this implies that all local multi-point correlation functions within subsystem $A$ can be computed as averages calculated with $\rho_E$.
If a system thermalises, as expected to be the case for non-integrable models \cite{nonint,rdo-08,r-09,bch-11,bdkm-12,rs-12,sks-13,r-14}, 
$\rho_E$  is the Gibbs distribution $\rho_E\propto e^{-\beta H}$.  
 In the case of integrable systems which do not thermalise, $\rho_E$  is given by  a generalised Gibbs 
ensemble (GGE)\cite{GGE}  where all the \textit{local} integrals of motion are taken into account,  although some different results  
have been found for  initial states of a specific class \cite{f-14,noGGE1,noGGE2,noGGE3,noGGE4}.

In the present case of  the transverse field Ising chain, both for quenches from the ground state  \cite{cef,cef-i,cef-ii} and from excited states   \cite{bkc-14},  it has been proved that the stationary state is described by GGE.  It has also been shown \cite{fe-13a} that the post-quench occupation number operators  $n_k=b_k^\dagger b_k$, although non-local quantities, can be written as linear combinations of the local integrals of motion. Thus the GGE density matrix constructed with local integrals of motion and the one constructed with $ n_k$ are equivalent, yielding
\begin{equation}
 \rho_{\rm GGE}=\frac{e^{-\sum_k{\lambda_k  n_k}}}{Z}\,,
\label{GGE}
\end{equation}
where  $\lambda_k$ are  fixed by matching the expectation values of the post-quench occupation number
 with their values in the initial state, i.e. imposing 
$ \langle  n_k\rangle_{\rm GGE}= \langle\Psi_0|  n_k|\Psi_0\rangle\equiv n_k$,
yielding \cite{bkc-14}
\begin{equation}
n_k=1-u_k^2m_k-v_k^2(1-m_{-k})\,,
\label{n_k}
\end{equation}
where 
\be
u_k=\cos(\Delta_k/2),\qquad v_k=\sin(\Delta_k/2)\,,
\label{uv}
\ee
and $\Delta_k$ is the difference between pre- and post-quench Bogoliubov angles having the 
explicit form in terms of $h$ and $h_0$  \cite{cef}
\be
\cos \Delta_k=\frac{h h_0- (h+h_0) \cos p_k+1}{\sqrt{1+h^2-2h\cos(p_k)} \sqrt{1+h_0^2-2h_0\cos(p_k)}}\,.
\ee
 Notice that the expectation values of the products of $n_k$ (which are also conserved) 
do not enter the GGE because of the cluster decomposition property of the considered  initial states \cite{sc-14} (see also \cite{cdeo-08}).

The other interesting thermodynamic ensemble is the so-called diagonal ensemble (DE) \cite{pol-11} 
\begin{equation}
\rho_\text{D}=\sum_j |c_j|^2|j\rangle\langle j|\,,
\end{equation}
where $c_j=\langle j|\Psi_0\rangle$ is the overlap of the eigenstate $|j\rangle$ of the post-quench Hamiltonian with the initial state. 
By definition, the DE captures, in the TD limit, the time averaged expectation values of \textit{all} observables, including non-local 
and non-stationary (e.g. oscillating) ones, irrespectively of the integrability of the system. 
Indeed, such an ensemble retains all the information about the initial state rather than a limited set of integrals of motion and, in 
this sense, is genuinely different from the canonical and generalized Gibbs ensembles since it has no relation with the 
economy of the maximum entropy principle.

\textbf{Inequivalence of the diagonal and the GGE entropy}.
The inequivalence of the diagonal ensemble and the GGE is captured by the difference of their entropies
\begin{align}
S_\text{D}&=-\mathrm{Tr}\rho_\text{D}\ln\rho_\text{D} = -\sum_j |c_j|^2 \ln |c_j|^2\,, \\
S_{\text{GGE}}&= -\mathrm{Tr}\rho_{\text{GGE}}\ln\rho_{\text{GGE}}\,.
\end{align}
reflecting the fact that there is some information loss in passing from the former to the latter.
Given that the GGE describes all local observables in a subsystem, 
the GGE density matrix coincides with the reduced density matrix of the subsystem.  
Clearly, the GGE entropy must then coincide with the extensive part of the long time limit of the entanglement entropy of a block of consecutive spins. 

As already mentioned, in quenches starting from the ground state in the transverse field Ising model \cite{g-13} and in the Lieb--Liniger model \cite{ckc-13} it was found that the GGE entropy is exactly twice the diagonal entropy. Here we set out to compare the various entropies for quenches starting from arbitrary excited states in the Ising model.

The vacuum state of the pre-quench Hamiltonian can be written in terms of the post-quench vacuum $|0\rangle$  
and mode creation operators as
\be
|0\rangle' = \prod_{k>0}(u_k-iv_k b_k^\dagger b_{-k}^\dagger)|0\rangle\,,
\ee
where $u_k$ and $v_k$ are given in Eq. (\ref{uv}).
It is straightforward to see that this state is annihilated by the pre-quench annihilation operator 
$b'_k =  u_k b_k + iv_k b^+_{-k}$ and it is normalised to 1. 

The excited initial state (\ref{ES}) can now be obtained by acting on the vacuum with the pre-quench creation operators written again in terms of $b_k,b^\dag_k.$ One finds
\begin{alignat}{4}
|\Psi_0\rangle &=\prod_{k>0}&&\left[(u_k-i v_k b_k^\dagger b_{-k}^\dagger)(1-m_k)(1-m_{-k}) \right. \nonumber\\ 
&& &+ \left. m_k (1-m_{-k}) b_k^\dagger + m_{-k} (1-m_{k}) b_{-k}^\dagger 
 \right. \nonumber\\ 
 &&&+ \left. m_k m_{-k} (u_k b_k^\dagger b_{-k}^\dagger -i v_k) \right]|0\rangle  =\nonumber\\
 &= \prod_{k>0}&&\left[\alpha _k  b_k^\dagger b_{-k}^\dagger + \beta _k +\gamma _k  b_k^\dagger +\delta _k  b_{-k}^\dagger   \right]|0\rangle\,,
\end{alignat}
where
\begin{align}
|\alpha _k|^2&=m_k m_{-k}+v_k^2(1-m_k-m_{-k})\,, \nonumber \\ 
|\beta _k|^2&=m_k m_{-k}+u_k^2(1-m_k-m_{-k})\,, \nonumber\\
 |\gamma _k|^2&=m_k(1-m_{-k})\,,\qquad |\delta _k|^2=m_{-k}(1-m_{k})\,.
\end{align}
The product over positive momenta only  originates from the fact that the Bogoliubov rotation connecting pre- and post-quench operators couples modes with opposite momenta.
The entropy of the diagonal ensemble is then
\begin{multline}
\label{Sd}
S_\text{D}=-\sum _{k>0}\Big[|\alpha _k|^2 \ln |\alpha _k|^2+|\beta _k|^2 \ln |\beta _k|^2\\ 
+ |\gamma _k|^2\ln |\gamma _k|^2+|\delta _k|^2 \ln |\delta _k|^2  \Big]\,.
\end{multline}
On the other hand, the GGE  entropy  is
\be
\label{Sgge}
\begin{split}
S_{\mathrm{GGE}}&=
-\sum _k\left[n_k\ln n_k+(1-n_k)\ln (1-n_k)\right]=\\
&= -\sum _k\Big\{[1-u_k^2(1-m_k-m_{-k})-m_{-k}] 
\\ &\qquad  \;\; \times \ln [1-u_k^2(1-m_k-m_{-k})-m_{-k}] \\
&\qquad\;\;  +[m_{-k}+u_k^2(1-m_k-m_{-k})] \\ 
&\qquad  \qquad\times \ln [m_{-k}+u_k^2(1-m_k-m_{-k})]\Big\}\,,
\end{split}
\ee
where we expressed $n_k$ in terms of the pre-quench occupation numbers $m_k$ via Eq. (\ref{n_k}).

Note that $S_D$ is defined through a sum over only positive modes, in accordance with the BCS-like structure of the initial state; instead  $S_\text{GGE}$ involves a sum over all $k$,  in agreement with the fact that in the GGE
correlations between modes $k$ and $-k$ are absent.

Interestingly, the fact that $m_k\in\{0,1\}$ implies that the seemingly very different summands of Eqs. \eqref{Sd} and \eqref{Sgge} are equal. In fact, only the modes with $m_k=m_{-k}$ contribute and a simpler expression for the entropies can be written as
\begin{subequations}
\label{Sfin}
\begin{align}
S_\text{D}&=\sum_{k>0}\big[m_km_{-k}+(1-m_k)(1-m_{-k})\big] s_k\,,\label{Sdmicro}\\
S_\text{GGE}&=\sum_k\big[m_km_{-k}+(1-m_k)(1-m_{-k})\big] s_k\,,
\end{align}
\end{subequations}
where we introduced $s_k\equiv- (u_k^2\ln u_k^2+v_k^2\ln v_k^2).$
This means that the relation
\be
S_\text{GGE}=2 S_\text{D}
\label{law2}
\ee
holds even for excited initial states. 
The reason for this factor of $2$ can be understood in terms of the argument of Ref. \cite{g-13} according to which 
the $\{k,-k\}$ pairs generated by the initial state contribute to the diagonal entropy but generate correlations that are invisible to the GGE.
Indeed, they have no influence on the reduced density matrix of a finite subsystem $A$:  
if a particle with momentum $k$ is in $A$, for long enough time, the $-k$ partner is surely outside $A$ \cite{g-13}.

At this point it must be stressed that the above results for both diagonal and GGE entropies are strictly valid only for 
finite systems because in several points we used that $m_k\in \{0,1\}$. 
Taking the TD limit is a complicated matter because $m_k\to m(p)$ a function that can be different 
from $0$ and $1$ and the way how the average is taken is very important.  
To make clearer where the problems stands 
a simple example is given by the states in which every other momentum state is filled. 
If the state is symmetric with respect to $k\to-k$ the entropy is maximal $S_{GGE}=\sum s_k$
and if the state is antisymmetric then the entropy is zero.
But both these states in the TD limit converge to the same function $m(p)=1/2$. 
It is then clear that the TD value of the entropy is not univocally determined by the function $m(p)=1/2$
as it is the case for correlation functions \cite{bkc-14} and the way we take the averages very important.
For example, one could consider the following two averaged entropies
\begin{align}
S^{TD} &=\overline{ -{\rm Tr}  \rho \ln \rho}\,,\\
S^{ST} &= -{\rm Tr} \overline{ \rho} \ln \overline{\rho}\,,
\end{align}
where the bar stands for the average over the initial states.
The two averages give different results which will be explicitly carried out in the remainder of the paper.
However, by no means one is more fundamental than the other. 
Indeed $S^{TD}$ looks more like a TD quantity in which the average is performed directly over the observable,
while $S^{ST}$ is the entropy of the average state. 
Roughly speaking the two averages are like quenched and annealed disorder in random systems
and indeed  similar features have also been noticed for the entanglement entropies in random spin chains 
\cite{fmc-10,m-14}.

\textbf{Integral representation for $S^{TD}$ in the TD limit.}
To write an integral formula for the entropies in the TD limit is not a completely trivial exercise. The reason is that the formula should take into account the correlations between modes $k$ and $-k$. In particular, due to the appearance of both $m_k$ and $m_{-k}$ in Eq. \eqref{n_k}, the post-quench occupation number function $n_k$ is very far from being a continuous function even if a smooth $m(p)$ captures very well the density of filled pre-quench modes.

This means that by simply replacing the discrete $m_k$ with a continuous $m(p)$ characteristic function in the formulas \eqref{Sd} and \eqref{Sgge} gives two different but equally incorrect results. However, in the derivation of expressions \eqref{Sfin} we already used the $\{k,-k\}$ correlation and $m_k\in\{0,1\}$, and a similar replacement in these turns out to yield the correct result, as we show below.

A systematic way to turn the discrete sums \eqref{Sfin} into coarse grained integral expressions is the following. We break up the set of momenta into  pairs of intervals, $I_1=[p,p+\Delta p]$ and $I_2=[-p,-p-\Delta p],$ each containing many momenta but sufficiently small such that within each the $m(p)$ function can be regarded as constant. Both intervals have $M=N\Delta p/(2\pi)$ momentum slots, of which $M_1=m(p)N\Delta p/(2\pi)$ and $M_2=m(-p)N\Delta p/(2\pi)$ are filled. The term $s_k$ can be regarded as constant over the short intervals, so we only have to focus on
\be
\Sigma(\{m_k\})=\sum_{k\in I_1}\big[m_km_{-k}+(1-m_k)(1-m_{-k})\big]\,.
\ee
There are 
\be
\#\text{config.} = \binom{M}{M_1}\binom{M}{M_2}
\ee
different microscopic $\{m_k\}$ configurations  satisfying the constraint that the total occupations of the two intervals are $M_1$ and $M_2$. We compute the distribution of their contributions and show that in the TD limit it becomes sharply peaked around its mean value.

The sum $\Sigma(\{m\})$ for each configuration gives the number of momenta $k$ for which both $k$ and $-k$ modes are either filled or empty. We call these ``good momenta'' hereafter. Without loss of generality we can assume that $M_1\ge M_2$. The configurations can be grouped into classes: the $n$th class consists of configurations in which for exactly $n$ out of the $M_2$ momenta the opposite momentum state is one of the empty $N-M_1$ states. There are 
\be
\binom{M}{M_1}\binom{M_1}{M_2-n}\binom{M-M_1}n
\ee
such configurations and they have $(M_2-n)+(M-M_1-n)$ good momenta ($M_2-n$ filled and $M-M_1-n$ empty pairs). The average value of $\Sigma(\{m\})$ is 
\be
\overline\Sigma=\frac{\sum_\text{config.}\sum_{k\in I_1}\big[m_km_{-k}+(1-m_k)(1-m_{-k})\big]}{\#\text{config.}}\,,
\ee
where the sum over the configurations in the numerator is given by
\begin{multline}
\binom{M}{M_1}\sum_{n=0}^{M_2} \binom{M_1}{M_2-n}\binom{M-M_1}{n} (M-M_1+M_2-2n)   \\
=\binom{M}{M_1}\binom{M}{M_2} \frac{M_1 M_2 +(M-M_1)(M-M_2)}{M}
\end{multline}
which gives
\be
\begin{split}
\overline{\Sigma}&=M\left[\frac{M_1}M\frac{M_2}M + \left(1-\frac{M_1}M\right)\left(1-\frac{M_2}M\right)\right]\\
&=M\big(m(p)m(-p)+\left[1-m(p)\right]\left[1-m(-p)\right]\big)\,.
\end{split}
\ee
We can also calculate the variance of the distribution:
\be
\begin{split}
\sigma^2 &= \frac1{ \binom{M}{M_1}\binom{M}{M_2}}
\binom{M}{M_1}\sum_{n=0}^{M_2} \binom{M_1}{M_2-n}\binom{M-M_1}{n} \\ 
& \qquad  \times \left(M-M_1+M_2-2n-\overline\Sigma\right)^2   \\
&=\frac{4M_1 M_2 (M-M_1)(M-M_2)}{(M-1)M^2}\\ &=\frac{M}{1-M^{-1}}\,4m(p)m(-p)\left[1-m(p)\right]\left[1-m(-p)\right]\,.
\end{split}
\ee
In the TD limit the number of momenta in the interval $I$ goes to infinity while the function $m(p)$ is kept fixed, then $M$ scales with $N$ and the relative variance vanishes:
\be
\sigma/\overline\Sigma\to\sim\frac1{\sqrt{N}}\,.
\ee
One can even  show that the discrete distribution approaches the normal distribution $\mathcal{N}(\overline\Sigma,\sigma)$ in the TD limit.

Since the distribution is sharply peaked we can substitute the contribution of the interval $I$ to the entropy by its average $\overline\Sigma$. The remaining sum over the small momentum intervals  can then be straightforwardly written as an integral in the TD limit:
\begin{multline}
\frac{S_\text{GGE}}N =\\ - \int_{-\pi}^\pi \frac{dp}{2\pi}\big\{m(p)m(-p)+\left[1-m(p)\right]\left[1-m(-p)\right]\big\}
s(p)
\label{Scont}
\end{multline}
with $s(p)=s_{k(p)}$, which agrees with the naive substitution $m_k\to m(p)$ and $\sum_k\to N\int dp/(2\pi)$ in Eqs. \eqref{Sfin} but not in Eqs. \eqref{Sd} and \eqref{Sgge}. 

Naturally, this expression holds for the overwhelming majority of microscopic states that are described by the smooth $m(p)$ function, but one can always construct very atypical states for which the formula fails 
such as those reported above in which every other momentum state is filled. 
These rare states are of course also present in any kind of coarse grained statistical physics description.

We also checked through numerical experiments that for random microscopic states generated to follow a given $m(p)\sim e^{-p/2}$ characteristic function, the continuum formula \eqref{Scont} indeed agrees with the microscopic calculation of the entropy in Eqs. \eqref{Sfin}. 

\begin{figure}
\includegraphics[width=0.45\textwidth]{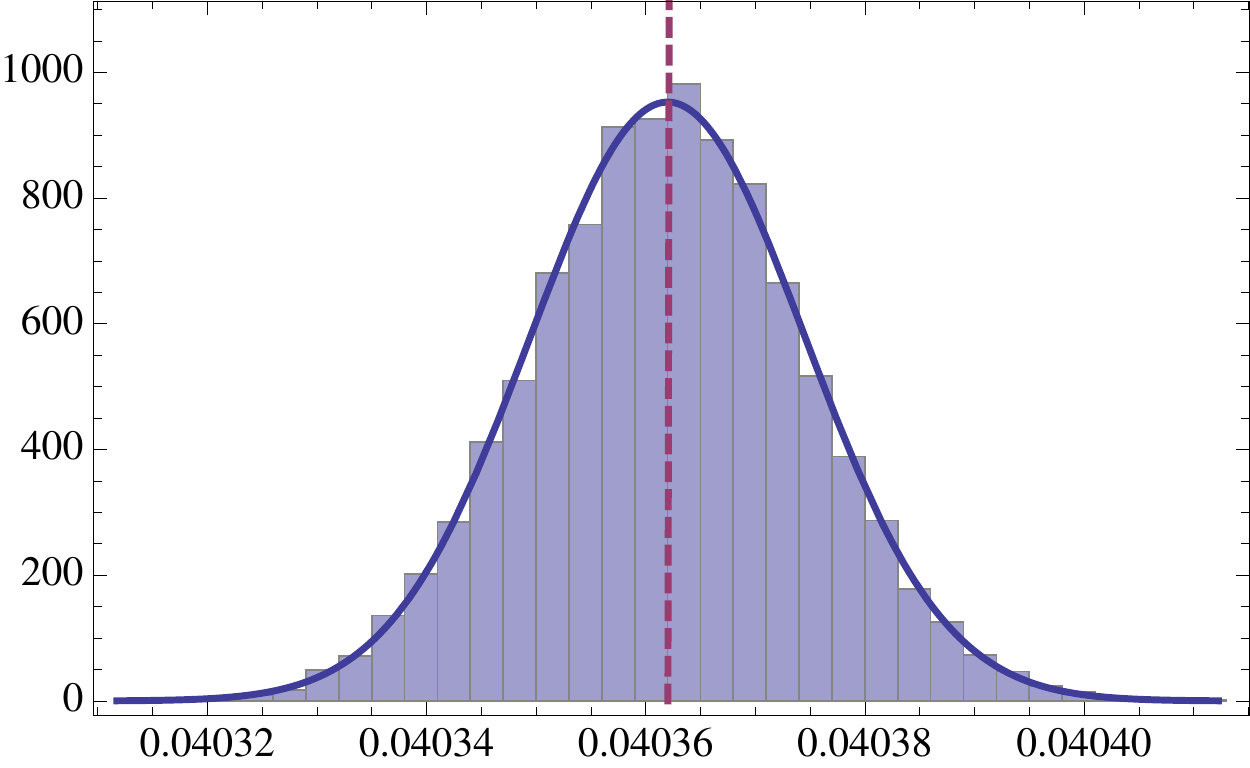}
\caption{Histogram of the diagonal entropy computed by Eq. \eqref{Sdmicro} using $10^4$ randomly generated microscopic $\{m_k\}$ configurations in which $\sim1000$ momenta are distributed as $\sim e^{-k/2}$  on a chain of length $L=5\times10^4$. The parameters of the quench are $h_0=7$, $h=2$. The dashed vertical line is the result of the continuum formula in Eq. \eqref{Scont}.}
\end{figure}

\textbf{Integral representation for $S^{ST}$ in the TD limit and comparison with the entanglement entropy}.
The calculation of $S^{ST}$ in the TD limit is instead much easier using the results of Ref. \cite{bkc-14}. 
Indeed, taking the average of the density matrix corresponds to averaging its elements which are multipoint correlation functions;  by Wick's theorem they can be deduced from the two-point fermionic functions which have 
already been calculated in Ref. \cite{bkc-14}. 
Just by plugging those results in the density matrix one obtains 
\begin{multline}
S^{ST}_\text{GGE} =  N \int_{-\pi}^{\pi} \frac{dp}{2\pi} H[m(-p)-m(p) +\\
(m(p)+m(-p)-1) \cos\Delta(p)],
\label{Sst}
\end{multline}
where $H(x)=-\frac{1+x}{2} \ln\left(\frac{1+x}{2}\right)-\frac{1-x}{2}\ln\left(\frac{1-x}{2}\right).$
Note that $S^{ST}_\text{GGE}$ is 
the naive continuum limit of $S_\text{GGE}$ in Eq. \eqref{Sgge} obtained by substituting $m_k\to m(p)$  
and replacing the sum with an integral. 
The result \eqref{Sst} is identical to the extensive part of  the stationary entanglement entropy calculated in Ref. \cite{bkc-14} in 
which indeed the average of the (time dependent) reduced density matrix has been considered.
This confirms that GGE and entanglement entropies are always equal when calculated consistently. 

\textbf{Conclusions}.
We computed the stationary entropies of the GGE and of the diagonal ensemble after a quench from excited states of the Ising model.
We have found that the GGE entropy is always twice the diagonal one, as already found for quenches starting  
from the ground state \cite{g-13}. 
It remains an interesting open issue to understand whether simple relations between GGE and diagonal entropies exist in interacting 
integrable models for which the initial states can be in principle a more complicated superposition of elementary 
excitations \cite{fm-10,sfm-12,ce-13,m-13,stm-13,nwbc-13,bse-14}. 
Furthermore, we showed that the thermodynamic limit of the entropies strongly depends on how the average over the initial 
micro-states with the same macroscopical features is taken, and which one is more relevant depends in principle on which real 
experiment (or numerical simulation) one aims to describe.

\textit{Acknowledgements}:
MK thanks G\'abor Tak\'acs for useful discussions. PC acknowledges the ERC  for financial  support under Starting Grant 279391 EDEQS. 
MK acknowledges financial support from the Marie Curie IIF Grant PIIF-GA-2012- 330076.

\end{document}